\documentclass[12pt]{article}

\begin{document}

\title{Existence of a Semiclassical Approximation in Loop Quantum Gravity}

\author{Marco Frasca \\
Via Erasmo Gattamelata, 3 \\
00176 Roma (Italy)}

\date{\today}

\maketitle

\begin{abstract}
We consider a spherical symmetric black hole in the Schwarzschild metric and apply Bohr-Sommerfeld
quantization to determine the energy levels. The canonical partition function is then computed and we show
that the entropy coincides with the Bekenstein-Hawking formula when the maximal number of states for
the black hole is the same as computed in loop quantum gravity, proving in this case the existence
of a semiclassical limit and obtaining an independent derivation of the Barbero-Immirzi parameter.
\end{abstract}


One of the most important open questions of loop quantum gravity \cite{rovelli} is the proof of the existence
of a semiclassical limit. This aspect is relevant to show that this approach to quantum gravity has indeed a
proper limit to a classical description of the gravity field.

Notwithstanding the missing of a proper semiclassical limit, there have been notable tentatives to study black
hole physics recurring to Bohr-Sommerfeld approximation. The most important of these applications is the so
called Bekenstein-Mukhanov effect \cite{bekmuk,kastrup} where, assuming a kind of quantization rule for the area of
the black hole, a quantization of its mass and then energy levels are obtained. The effect is to modify the
spectrum of the Hawking radiation producing identical envelope but discrete lines. So, energy levels are
predicted for a black hole but this effect does not appear to realize itself for loop quantum gravity due
to a different rule of quantization of area \cite{rovelli} and the Bekenstein-Mukhanov effect appears just
a semiclassical effect in the limit of large quantum numbers.Besides, it was pointed out recently
that the spectrum of Hawking radiation can appear discrete just in four dimensions \cite{das}.

In this paper we will work out an analysis of a Schwarzschild black hole in the same spirit but starting
from the solution of the Hamilton-Jacobi equations \cite{landau}. The result we obtain is striking as the
same counting of the maximal number of levels for the black hole is produced as in loop quantum gravity giving
a consistent support to both the approaches. As a by-product we gain a clue of the existence of the semiclassical
limit for loop quantum gravity ensuring an explicit expression for the wave function. The proof we give yields insights
on the properties of the propagator itself for this case. 

To start our analysis we consider the Kaplan solution of stable circular orbits in a Schwarzschild black hole\cite{kaplan}. One has the Hamilton-Jacobi equations in the Schwarzschild metric (here and in the following $c=1$)\cite{landau}
\begin{equation}
    \left(1-\frac{r_S}{r}\right)^{-1}\left(\frac{\partial S}{\partial t}\right)^2
    -\left(1-\frac{r_S}{r}\right)\left(\frac{\partial S}{\partial r}\right)^2
    -\frac{1}{r^2}\left(\frac{\partial S}{\partial\theta}\right)^2-M^2=0
\end{equation}
being $r_S=2GM$ the Schwarzschild radius, $M$ the mass of the black hole and $S$ the action. It is straightforward
to solve this equation by setting $S=-Et+L\theta+S_r(r)$ being $E$ the energy and $L$ the angular moment integration constants. The equation relevant for our aims is given by the condition $\partial S/\partial E=constant$ yielding
the equation
\begin{equation}
    \left(1-\frac{r_S}{r}\right)^{-1}\frac{dr}{dt}=\frac{1}{E}\sqrt{E^2-V(r)^2}
\end{equation}
being
\begin{equation}
    V(r) = M\left[\left(1-\frac{r_S}{r}\right)\left(1+\frac{L^2}{M^2r^2}\right)\right].
\end{equation}
Then, stable circular orbits are obtained for
\begin{equation}
    \frac{r}{r_S}=\frac{L^2}{M^2r_S^2}\left(1+\sqrt{1-\frac{3M^2r_S^2}{L^2}}\right).
\end{equation}
and energy
\begin{equation}
    E=L\sqrt{\frac{2}{rr_S}}\left(1-\frac{r_S}{r}\right).
\end{equation}

Application of Bohr-Sommerfeld quantization condition for circular orbits is straightforward in this case 
simply setting\cite{bojo} $L=n\hbar$ being $n$ an integer giving
\begin{equation}
    \frac{r_n}{r_S}=\frac{n^2\hbar^2}{M^2r_S^2}\left(1+\sqrt{1-\frac{3M^2r_S^2}{n^2\hbar^2}}\right).
\end{equation}
We want to analyze this equation for a physical macroscopic black hole. So, let us introduce
the Compton length of the test body $\lambda_c=\hbar/M$. This will be a very small number, so,
the first physical orbit will appear at $n_{lim}=\sqrt{3}r_S/\lambda_c$ that is a very large
number. But we are still near the horizon as $\frac{r_{n_{lim}}}{r_S}=3$. Then, a small increase
in $n$ beyond $n_{lim}$ will make negligible the second term inside the square root but
it will not take us too far from the black hole horizon. This means that we are really
sensing the behavior of test particles in a semiclassical regime as we want.

Then, due to the physics of the problem,
the second term inside the square root can be neglected, leaving just the Newtonian limit
\begin{equation}
    \frac{r_n}{r_S}\approx\frac{2n^2\hbar^2}{M^2r_S^2}
\end{equation}
that is
\begin{equation}
    r_n\approx\frac{n^2\hbar^2}{GM^3}.
\end{equation}
One has immediately the energy levels in the same approximation
\begin{equation}
    E_n\approx M-\frac{2G^2M^5}{n^2\hbar^2}
\end{equation}
giving energy levels for the black hole in the form expected for a Newtonian potential. 
In the following we will omit the term $M$ in the energy being just a constant added. 
Besides, we will take as irrelevant the contribution of the continuous part of the spectrum 
to the computation of black hole entropy.

The canonical partition function should be straightforward to be written down as
\begin{equation}
    Z=\sum_{n=1}^\infty e^{\beta\frac{E_0}{n^2}}
\end{equation}
being $E_0=\frac{2G^2M^5}{\hbar^2}$, and $\beta=1/kT$ with $k$ the Boltzmann constant and $T$ 
the temperature, but this series gives $\infty$ and then the partition
function does not exist. This problem is known since long time and was faced firstly by 
Fermi \cite{fermi} that introduced a convergence factor into the series due to the limited
number of levels entering into a finite volume. This convergence factor can be properly
computed by the propagator of a particle in a $1/r^2$ potential as was done by Blinder
\cite{blinder} and gives us the converging series
\begin{equation}
    Z=\sum_{n=1}^\infty e^{\beta\frac{E_0}{n^2}}e^{-\frac{\pi}{4}\frac{n^2}{N^2}}
\end{equation}
being $N$ a parameter to be computed and will be our aim to obtain it 
for a Schwarzschild black hole. We note that $N$ is the maximal value of the quantum
number $n$ given by the physics of the problem. Then, it is straightforward to verify that,
in the semiclassical limit where $N$ should be very large, the series is fairly well approximated by
\begin{equation}
\label{eq:nM}
    Z\approx N
\end{equation}
meaning for the free energy
\begin{equation}
    F\approx-kT\ln\left(N\right).
\end{equation}
Entropy is given by
\begin{equation}
    S=-\frac{\partial F}{\partial T} = k\ln\left(N\right).
\end{equation}
We note as $N$ defines the
number of states needed to define the entropy of a gas of particles orbiting near the horizon
of the black hole where the quantum nature of the gravitational field can be sensed
and we face the semiclassical regime. Assuming
equilibrium between such a gas of test particles and the black hole we can 
equate this expression to the Bekenstein-Hawking formula
$S=kA/4G\hbar$, being $A$ the horizon area of the black hole, and one has immediately
\begin{equation}
    N=2^{\frac{A}{4G\hbar\ln 2}}.
\end{equation}
We recognize the same formula for the counting of states of a black hole in loop quantum gravity 
when the Barbero-Immirzi paramter is $\gamma = \ln 2/\pi\sqrt{3}$ \cite{rovelli,ash1,ash2,ash3}. It can be
seen in this way that the argument is fully consistent turning out $N$ very large so to have
both the semiclassical approximation and eq.(\ref{eq:nM}) working. This maximal number
of states warrants a finite entropy for a gas of test particles near the horizon of a black
hole and is strictly related to the physical properties of the black hole itself.

At this stage we have proved that, in a different way, the counting into the entropy of a black
hole turns out to be the same as the one computed in loop quantum gravity \cite{rovelli}. But we
have worked in a complete semiclassical setting enforcing the value of the Barbero-Immirzi
parameter as computed by \cite{ash1,ash2,ash3} in disagreement with the recent analysis \cite{meiss,dl}. This latter
analysis was criticized by \cite{alex} supporting the semiclassical limit we have exploited here. 
A nice account of the situation, linking it to the normal modes of a black hole, 
is given in \cite{ns}. 

The relevance of our computation resides in the derivation, in another way, of the counting
of black hole states proved to be the same as in the initial computation done in 
loop quantum gravity\cite{ash1,ash2,ash3}. Both quantization
of angular momentum in the semiclassical limit
and equilibrium between orbiting test particles and the black hole near
horizon were needed hypothesis.

\end{document}